\documentclass[12pt]{article}
\newcommand{\be}{\begin{equation}}
\newcommand{\ee}{\end{equation}}
\newcommand{\bdm}{\begin{displaymath}}
\newcommand{\edm}{\end{displaymath}}
\newcommand{\bea}{\begin{eqnarray}}
\newcommand{\eea}{\end{eqnarray}}

\begin{document}
\begin{center}
{\Large\bf  Rescattering effects and  the $\sigma$ pole  in hadronic  decays}
\\

I.~Caprini\\ 

National Institute of Physics and Nuclear
  Engineering,\\  R-077125,  Bucharest, Romania
\vskip0.5cm

{\bf Abstract}\end{center}
The $\sigma$ resonance was observed  as a conspicuous $\pi^+\pi^-$ peak in hadronic decays like $J/\psi\to \pi^+\pi^-\omega$ or
 $D^+\to\pi^+\pi^-\pi^+$. The phase of the  $\sigma\to\pi^+\pi^-$  amplitude, extracted from production data  within the conventional isobar model, is assumed to coincide with that  in  $\pi\pi$ elastic scattering. We check the validity of this assumption by using Lehmann-Symanzik-Zimmermann (LSZ) reduction and unitarity. The rescattering effects in the final three-particle states  are shown to generate a correction to the phase given by a naive application of Watson theorem. We briefly discuss the implications of this result for the  pole  determination from production data.
%%%%%%%%%%%%%%%%%%%%%%%%%%%%%%%%%%%%%%%%%%%%%%%%%%%%%%%%%%%
\section{Introduction}
The lowest scalar resonance $\sigma$ (or $f_0(600)$) appears as a pole on the second Riemann sheet of the  $I=l=0$ partial wave amplitude of $\pi\pi$ elastic scattering (we denote this wave as $t_0^0(s)$). Although a typical  resonant behaviour  is not seen, because the pole is far from the real axis and  is compensated  in the physical region by the Adler zero, many determinations of the sigma pole are based on  $\pi\pi$ scattering \cite{PDG 2004}. However,
  the pole was usually extracted from parametrizations valid along the physical region. The predictions are therefore  affected by the large uncertainties of the analytic extrapolation to a distant point. Recently, a  model-independent extrapolation into the complex plane, based on the Roy equation for $t_0^0(s)$, led  to a precise prediction of the pole position \cite{CCL}. 

The $\sigma$ resonance was also seen  as a  peak in  BES II data on  $J/\psi\to \pi^+\pi^-\omega$ \cite{BES} and  in the data on
 $D^+\to\pi^+\pi^-\pi^+$ reported by E791 Collaboration \cite{E791}. The conspicuous sign in production processes  is explained by the absence of the  Adler zero \cite{Bugg, Oller}.  It is of interest to compare  the pole determinations from production processes and $\pi\pi$ elastic scattering.  In the present paper we consider some issues related to this problem.

   To illustrate the discussion we consider  the strong decay 
\be\label{J}
J/\psi(p)\to\pi^+(p_1)+\pi^-(p_2)+\omega(p_3)\,,
\ee
 but our arguments apply also to the decay $
D^+\to\pi^+\pi^-\pi^+$,
and more generally to $h\to\pi^+\pi^-h_1$,
where  $h$ and $h_1$ are hadrons.
  We define the Mandelstam variables 
\be\label{Mand}
s=(p_1+p_2)^2,\quad t=(p_1+p_3)^2,\quad u=(p_2+p_3)^2,
\ee
which satisfy $s+t+u=m_{J/\psi}^2+2 m_\pi^2+m_\omega^2$.
The physical region of the process (\ref{J}) corresponds to  $s>4 m_\pi^2$, $t>(m_\pi+m_\omega)^2$ and $u>(m_\pi+m_\omega)^2$.
 Since some particles have nonzero spins, a decomposition in Lorentz covariants is required.  For simplicity, we neglect this complication and consider an invariant amplitude  $A(s,t)$ as a function of the Mandelstam variables (\ref{Mand}). 

In the conventional isobar model,  the amplitude of the decay  (\ref{J}) is expressed as a sum of isobaric resonances in various channels\footnote{A complex  constant accounting for the direct nonresonant interaction is sometimes added to the resonances (see \cite{E791, Oller}).}. In a diagrammatic language, the three-body decay is assumed to be described by tree diagrams where the
 production of two-body final states proceeds via intermediate resonances.  More exactly,  the amplitude is written as \cite{BES, Bugg}
\be\label{IM}
 A(s,t)= A_s(s,t)+ [A_t(t, s)+ (t\leftrightarrow u)]\,,
\ee
where  the $s$-channel and $t$ ($u$)-channel amplitudes  are expanded as
\be\label{Ast}
 A_s(s, t)=\sum a_l(s) P_l(\cos\theta_s)\,,\quad
  A_t(t, s)=\sum b_l(t) P_l(\cos\theta_t)\,.
\ee
In these relations
 $\theta_s(=\theta_{13})$ is  the angle between the three-momenta of $\pi(p_1)$  and  $\omega(p_3)$ in the rest system of the two pions, and $\theta_t(=\theta_{12})$  the angle between the three-momenta of $\pi(p_1)$  and  $\pi(p_2)$ in the rest system of  $\pi(p_1)\omega$. The lowest partial waves in (\ref{Ast}) are assumed to be dominated by resonances.  For the process (\ref{J}) the $s$-channel resonances $\sigma$, $f_0(980)$ and $f_2(1270)$ contribute to the partial waves $a_0$ and $a_2$, respectively, and $b^+_1(1235)$ appears in both $S$ and $D$-waves of the $t$ ($u$) channels. Keeping for simplicity only the contributions of $\sigma$ and $b_1$ and assuming  Breit-Wigner parametrizations, one writes \cite{Bugg}  
\be\label{a0b0}
 a_0(s)=\frac{C_\sigma {\rm e}^{i\Delta_\sigma}}{m_\sigma^2 -s-i m_\sigma \Gamma_\sigma(s)}\,,\quad\quad 
b_0(t)=\frac{C_{b_1} {\rm e}^{i\Delta_{b_1}} }{m_{b_1}^2-t -i m_{b_1} \Gamma_{b_1}}\,.
\ee
 In Refs. \cite{BES, Bugg}, $a_0(s)$ is denoted as the $\sigma\to\pi\pi$ amplitude.

The  phases $\Delta_\sigma$ and $\Delta_{b_1}$ appearing in (\ref{a0b0}) account for the interactions of $\sigma\omega$ and $\pi b_1$, respectively. In  the conventional isobar model \cite{BES, E791}, these phases are assumed to be independent on the Mandelstam variables.
 Moreover, by invoking Watson theorem \cite{Watson}, 
  the phase of the $\sigma\to\pi\pi$ amplitude $a_0(s)$  was assumed \cite{Bugg, Oller} to coincide,  up to a constant, with the pion-pion phase shift $\delta_0^0$ appearing in the expression of the $l=I=0$ partial wave:
\be\label{t00} 
t_0^0(s)= \frac{1}{2 i \rho(s)} \{\eta^0_0(s)  e^{2 i \delta_0^0(s)}-1\} \,,
\ee
where $\rho(s)=\sqrt{1- 4m_\pi^2/s}$. An equivalent formulation  is to assume \cite{Oller} that  the denominator of the function $a_0(s)$ given in (\ref{a0b0})  coincides with the function  $D(s)$, appearing in the $N/D$ method \cite{MartinSpearman} for calculating the amplitude $t^0_0(s)$.

 The purpose of this letter is to check  the validity of  Watson theorem in the isobar model for decay processes.  We recall that  the theoretical difficulties of the three-particle decays are known since a long time. Anomalous singularities generated by rescattering effects and three-body dispersion relations were considered by several authors (see 
\cite{Anis1, Anis2} and older references quoted therein). In the present paper we investigate the
 phases of the amplitudes defined in the isobar model, using an approach based on LSZ reduction and unitarity.

%%%%%%%%%%%%%%%%%%%%%%%%%%%%%%%%%%%%%%%%%%%%%%%%%%%%%%%
\section{LSZ reduction and unitarity}
We start from the $S$-matrix element of the process (\ref{J})
\begin{equation}\label{defS}
S_{fi}=\langle \pi(p_1)\, \pi(p_2) \omega(p_3); {\rm out\,}| J/\psi(p); {\rm in}\rangle\,.
\end{equation}
After  the LSZ reduction \cite{LSZ, Bart} of the $\omega$
meson we obtain
\be\label{Sfi}
S_{fi}=\delta_{fi}+\frac{i}{\sqrt{2 p_{3,0}}}\int {\rm d}x e^{i p_3\cdot x}\langle \pi(p_1)\, \pi(p_2) 
; {\rm out\,}|\eta_\omega(x)| J/\psi(p); {\rm in}\rangle\,,
\ee
where  $p_{3,0}$ is the time component of $p_3$ and $\eta_\omega={\cal K}_x \phi_{\omega}(x)$ denotes the source operator (${\cal K}_x $ is the Klein-Gordon operator and $\phi_{\omega}$ the
interpolating field of the omega meson).
In what follows we do not need the explicit 
expressions of the sources, but only the significance of the matrix elements 
involving them.

Using translational invariance $
\eta_\omega(x)=e^{iP\cdot x} \eta_\omega(0) e^{-iP\cdot x}$
where $P$ denote the momentum operator, we write (\ref{Sfi}) as
\be
S_{fi}=\delta_{fi}+ \frac{i}{\sqrt{2 p_{3,0}}} (2\pi)^4 \delta(p_1+p_2+p_3-p)\langle \pi(p_1)\, \pi(p_2) 
; {\rm out\,}|\eta_\omega(0)| J/\psi(p); {\rm in}\rangle\,.
\ee
From the general expression of the $S$-matrix in terms of the invariant
 amplitude $A(s,t)$, it follows that 
\be\label{LSZA}
A(s,t) = \frac{{\cal N} }{\sqrt{2 p_{3,0}}}\langle \pi(p_1)\, \pi(p_2) 
; {\rm out\,}|\eta_\omega(0)| J/\psi(p) {\rm in}\rangle\,.
\ee
where ${\cal N}= 4\sqrt{p_{0}  p_{1,0}  p_{2,0} p_{3,0}}$ is a normalization factor. In the same way we express the 
invariant amplitude  $T(s,t)$ of the elastic scattering  $\pi(k_1)+\pi(k_2)\to \pi(p_1)+\pi(p_2) $ as
\be\label{Tst}
 T(s,t') =  \frac{{\cal N}'}{\sqrt{2 p_{2,0}}}\langle \pi(p_1)\,|\eta_{\pi_2}(0)| \pi(k_1)\pi(k_2); {\rm in}\rangle\,,
\ee
where ${\cal N}'= 4\sqrt{p_{1,0} p_{2,0} k_{1,0}  k_{2,0}}$, and   the physical domain in the $s$ channel is defined by $s=(p_1+p_2)^2=(k_1+k_2)^2 >4 m_\pi^2$, $t'=(p_1-k_1)^2<0$.

By applying once more the LSZ reduction to the matrix element (\ref{LSZA}), we obtain:
\be\label{Ast1}
A(s,t)=  \frac{{\cal N} i } {\sqrt{2 p_{3,0}2 p_{2,0} }}\int {\rm d}x e^{i p_2\cdot x} \theta(x_0)\langle \pi(p_1)
|[\eta_{\pi_2}(x),\eta_\omega(0)]| J/\psi(p);{\rm in}\rangle\,,
\ee
where $\eta_{\pi_2}(x)$ is the source of the final pion $\pi(p_2)$. 

As it is known, the LSZ formalism allows the analytic continuation of  the amplitude 
$A(s,t)$ in the complex planes of the Mandelstam variables, where the expression (\ref{Ast1})  defines a holomorphic function (an important ingredient in the proof is causality, {\em i.e.} the fact that the retarded commutator vanishes for spacelike values $x^2<0$).
In what follows we  only use the LSZ representation
to derive the unitarity relation and explore its consequences.

By inserting a complete set of states $|n\rangle$ in the two terms of  the retarded commutator, 
 the matrix element appearing in (\ref{Ast1}) writes as
\be\label{sumn}
 \sum_{n} \left[\langle \pi(p_1)|\eta_{\pi_2}(x)|n \rangle  \langle n|\eta_\omega(0)| J/\psi\rangle -
 \langle \pi(p_1)|\eta_\omega(0)|n \rangle \langle
n| \eta_{\pi_2}(x)| J/\psi\rangle\right]\,.
\ee
 In the two particle approximation, the lowest states which contribute in the first sum are
 $|n\rangle=|\pi(k_1) \pi(k_2)\rangle$, where the two pions have $I=0$, while in the second sum 
 $|n\rangle=|\pi(k_1) \omega(k_2)\rangle$.  The sum over intermediate states involves integrations upon the momenta of the on-shell particles and sums over polarizations. After imposing the   translation invariance, we obtain from (\ref{Ast1}):
\bea\label{lsz}
 A(s,t)=i {\cal N} \int{\rm d}x \theta(x_0) \int \frac{{\bf{\rm d} k_1}}{(2\pi)^3} \frac{{\bf{\rm d} k_2}}{(2\pi)^3}e^{i p_2\cdot x+i p_1\cdot x -i k_1\cdot x -i k_2\cdot x} \nonumber\\
\times  \frac{\langle \pi(p_1)|\eta_{\pi_2}(0)|\pi(k_1)\pi(k_2)}{\sqrt{2 p_{2,0}}}\frac{\rangle  \langle   \pi(k_1)\pi(k_2) |\eta_\omega(0)| J/\psi(p)\rangle}{\sqrt{2 p_{3,0}}} \nonumber \\
- i {\cal N} \int{\rm d}x \theta(x_0)\int \frac{{\bf{\rm d} k_1}}{(2\pi)^3} \frac{{\bf{\rm d} k_2}}{(2\pi)^3}e^{i p_2\cdot x-i p \cdot x +i k_1\cdot x +i k_2\cdot x} \nonumber\\
\times  \frac{\langle \pi(p_1)|\eta_\omega(0)|\pi(k_1)\omega(k_2)}{\sqrt{2 p_{3,0}}}\frac{\rangle  \langle   \pi(k_1)\omega(k_2) |\eta_{\pi_2}(0)| J/\psi(p)\rangle}{\sqrt{2 p_{2,0}}}.
\eea
The presence of $\theta(x_0)$ in the integral of Eq. (\ref{lsz}) leads to discontinuities of the  amplitude across the real axis of the Mandelstam variables \cite{LSZ, Bart}. According to the general prescription, the discontinuity  is obtained formally from the
expression (\ref{lsz}) through the replacement of $i\theta(x_0)$ by $1/2$ \cite{Bart}. Then the integral upon $x$ gives $(2\pi)^4 \delta(p_1+p_2-k_1-k_2)$ 
 in the first term, and  $(2\pi)^4 \delta(p-p_2-k_1-k_2)$ in the second.
 Moreover, in the first term of (\ref{lsz})  we  recognize the product of the  amplitudes of the processes $\pi(k_1)+\pi(k_2)\to \pi(p_1)+\pi(p_2)$ and $J/\psi(p)\to\pi(k_1)+\pi(k_2)+\omega(p_3)$, while in the second term appears the product of the  amplitudes of the processes $\pi(k_1)+\omega(k_2)\to \pi(p_1)+\omega(p_3)$ and $J/\psi(p)\to\pi(p_2)+\pi(k_1)+\omega(k_2)$. 

From this discussion, it follows that  the amplitude  $A(s,t)$ has branch cuts for $s>4 m_\pi^2$ and $t>(m_\pi+m_\omega)^2$. By reducing the pion $\pi(p_1)$ instead of  $\pi(p_2)$ one obtains the branch cut for  $u>(m_\pi+m_\omega)^2$.  Hence, in the physical region of the decay (\ref{J}) all the variables $s$, $t$ and $u$ are above the unitarity thresholds. Singularities in both the $s$ and $t$ variables are expected to occur in each of the two terms in (\ref{lsz}). We recall that  in the isobar model defined in (\ref{IM})-(\ref{Ast}), the term $A_s(s,t)$ has singularities   only in the $s$ variable, being holomorphic with respect to $t$, while $A_t(t,s)$  has singularities  only in  $t$, being regular with respect to $s$. This shows the limitation  of the isobar model.

 For definiteness we consider that the complete set of intermediate states are ``out'' states and recall that for one particle states the ``in'' and ``out'' sets are equivalent. Recalling the definitions (\ref{LSZA}) and (\ref{Tst}) of the invariant amplitudes and focusing on the first term in (\ref{lsz}), responsible for the discontinuity with respect to the variable $s$,  we obtain 
\be\label{unit}
 \frac{1}{2 i}\{ A(s+i\epsilon, t)-A(s-i\epsilon, t)\} = \frac{1}{8 \pi^2 }\int \frac{{\bf{\rm d} k_1}}{2 k_{1,0}} \frac{{\bf{\rm d} k_2}}{2 k_{2,0}}\,\delta(P) \,T^*(s, t') A(s, t'')\,,
\ee
where $P=p_1+p_2-k_1-k_2$. The amplitudes are evaluated for $s=(p_1+p_2)^2=(k_1+k_2)^2$ and the momentum transfers $t'=(p_1-k_1)^2$ and $t''=(k_1+p_3)^2$, respectively. The integral (\ref{unit}) is easily evaluated in the c.m.s. of the two pions ${\bf p_1+p_2= k_1+k_2=0}$. After the trivial integrations due to the delta functions, (\ref{unit}) reduces to an integral   upon the angular variables:
\be\label{unit1}
\frac{1}{2 i}\{A(s+i\epsilon, t)-A(s-i\epsilon, t)\}= \frac{1}{64 \pi^2}\,\int {\rm d}\Omega \,\rho(s) T^*(s, t') A(s, t'')\,,
\ee
where ${\rm d}\Omega={\rm d} \phi\,{\rm d} \cos \theta''$,  $\theta''$ being the  angle  between  the three momenta of $\pi(k_1)$ and $\omega(p_3)$  in the pion rest system.  For the $\pi\pi$ isoscalar amplitude $T(s,t')$ we use the Legendre expansion
\be\label{pw}
T(s, t')= 16 \pi \sum\limits_{l=0}^\infty (2 l+1)  P_l(\cos \theta')\,t_l^0(s)\,, \ee
where $\theta'$ is the angle between the three momenta $p_1$ and $k_1$. We stress that (\ref{unit1}) is a general unitarity relation, independent of the isobar model. Let us restrict now to this model, taking for
 the amplitude  $A(s, t'')$ in (\ref{unit1}) the expression given in (\ref{IM})-(\ref{Ast}).
  Using the well known relation \cite{MartinSpearman}
\be\label{cos}
\cos\theta'' = \cos\theta \cos\theta'+ \cos \phi  \sin\theta \sin\theta'\,,
\ee 
the integral upon the angle $\phi$ is trivial. Recalling that only the first term  in (\ref{IM}) has a discontinuity for $s>4 m_\pi^2$, projecting onto the $S$-wave and using the orthogonality of Legendre polynomials we obtain
\bea\label{unit2}
\frac{1}{2 i}\{  a_0(s+i\epsilon)-a_0(s-i\epsilon)\}= \rho(s)\,(t_0^0(s))^* a_0(s)\nonumber\\
+\frac{\rho(s)}{2}(t_0^0(s))^* \int\limits_{-1}^{1} {\rm d}\cos \theta''[A_t(t'',s)+A_u(u'', s)].
\eea We recall that time reversal invariance implies the reality relation $a_0(s-i\epsilon)=a_0^*(s+i\epsilon)$, from which it follows that the l.h.s. of (\ref{unit2}) is real and equal to ${\rm Im}\, a_0(s)$ (if not otherwise specified,   $s$ is taken on the upper edge of the cut).
%%%%%%%%%%%%%%%%%%%%%%%%%%%%%%%%%%%%%%%%%%%%%%%%%%%%%%%%%%%%%%%%%%%%%%%%%
\section{ Watson theorem}
%%%%%%%%%%%%%%%%%%%%%%%%%%%%%%%%%%%%%%%%%%%%%%%%%%%%%%%%%%%%%%%%%%%%%%%%%
 Neglecting the four-pion channel which opens very slowly, the elastic region extends up to  the threshold for $K\bar K$ creation. Below this threshold $\eta_0^0(s)=1$, therefore the amplitude (\ref{t00}) becomes $t_0^0=e^{i \delta_0^0}\sin\delta_0^0/\rho(s)$. If we neglect the second term in the r.h.s. of (\ref{unit2}) we obtain
\be\label{unit3}
\frac{1}{2 i}\{  a_0(s+i\epsilon)-a_0(s-i\epsilon)\}= {\rm e}^{-i \delta_0^0(s)} \sin \delta_0^0(s)\ a_0(s)\,.
\ee
  This relation implies $a_0(s+i\epsilon)=a_0(s- i\epsilon) e^{2 i\delta_0^0(s)}$, which is equivalent to Watson theorem:
 the phase of $a_0(s)$ is equal (modulo $\pm \pi$)   to the phase shift  $\delta_0^0$. Alternatively, writing \cite{MartinSpearman}
\be\label{NoverD}
t_0^0(s)=\frac{N(s)}{D(s)}\,,
\ee
where $N(s)$ has only a left hand cut for $s<0$  and $D(s)$ a right hand cut for $s>4 m_\pi^2$,  a solution of (\ref{unit3}) has the form: 
\be\label{fNoverD}
a_0(s)=\frac{C(s)}{ D(s)}\,,
\ee
 where the function $C(s)$, real  for $s>4 m_\pi^2$, is arbitrary. By the uniqueness of analytic continuation, 
this implies that $t_0^0(s)$ and the function $a_0(s)$  have the same poles 
on the second sheet. We will come back on this point in the next section.

If the second term    in the r.h.s. of (\ref{unit2}) is not neglected we obtain, instead of (\ref{unit3}), the more general relation
\be\label{unit4}
{\rm Im}\, a_0(s)= {\rm e}^{-i \delta_0^0(s)} \sin \delta_0^0(s) [a_0(s) +  h(s)]\,,
\ee
where
\be\label{h}
h(s)=\frac{1}{2} \int\limits_{-1}^1 {\rm d}\cos\theta'' [A_t(t'',s)+ [A_u(u'',s)].
\ee
 We recall that the angles $\theta_s$ and $\theta_t$ in the expansions (\ref{Ast}) are expressed in terms of the Mandelstam variables, for instance:
\be\label{thetas}
 \cos \theta_s=\frac{m_\omega^2+m_\pi^2 + \sqrt{|p|^2+m_\omega^2}\,\sqrt{s}-t}{ 2 |p| \sqrt{s/4-m_\pi^2}}\,,
\ee
 where $|p|= \lambda^{1/2}(s, m_{J/\psi}^2, m_\omega^2)/(2 \sqrt{s})$ is the three momentum of $J/\psi$ ($\omega$) in the rest system of the pions (here $\lambda(a,b,c)=a^2+b^2+c^2-2 a b-2 ac-2 bc$). Using these relations and retaining in the expansion (\ref{Ast}) of $A_t+A_u$ only  the $S$-wave $b_0$, parametrized as in (\ref{a0b0}),  we have:
\be\label{h1}
h(s)\sim \frac{C_{b_1}\,  e^{i \Delta_{b_1}}} {2|p| |k_2|}  \ln\left[1+\frac{4 |p||k_2| }{m_{b_1}^2- m_\omega^2-m_\pi^2 -i \Gamma_{b_1}m_{b_1} - p_{3,0} \sqrt{s}- 2 |p||k_2| }\right]\,,
\ee
where $|k_2|=\sqrt{s/4-m_\pi^2}$ and $p_{3,0}= \sqrt{|p|^2+m_\omega^2}$, with  $|p|$ defined below (\ref{thetas}). If the isobar model contains also a  nonresonant term \cite{E791, Oller}, the function $h(s)$ will include its  contribution. 

From Eq. (\ref{unit4}) we can calculate a correction to Watson theorem. To this end  we impose time reversal invariance, which means that the r.h.s. of (\ref{unit4}) must be real. By requiring that the imaginary part vanishes, we obtain
\be\label{deltaPhi}
\sin[\Phi(s)-\delta_0^0(s)]=-\frac{{\rm Im}\left[ e^{-i\delta_0^0(s)} h(s)\right]}{ |a_0(s)|}\,,
\ee where   $\Phi(s)$ is the phase of the production amplitude:
\be\label{Phi}
a_0(s)=|a_0(s)| e^{i \Phi(s)}.
\ee
The relation (\ref{deltaPhi}) gives a calculable correction to the phase predicted by the naive application of Watson theorem in the elastic region. Above the inelastic threshold $s=4 m_K^2$ the elasticity $\eta_0^0(s)$ in (\ref{t00}) drops very quickly below unity,  and additional terms due to the $K \bar K$ intermediate states 
appear  in the r.h.s. of (\ref{unit2}). Since the unitarity sums for the scattering and the decay processes contain different contributions,  the phase of  $a_0(s)$   in the inelastic region may be quite different from  $\delta_0^0(s)$.

%%%%%%%%%%%%%%%%%%%%%%%%%%%%%%%%%%%%%%%%%%%%%
\section{Comments}
The above analysis shows that   the phase of the $\sigma\to\pi\pi$ amplitude $a_0(s)$, defined in  the conventional isobar model for  hadronic decays, is not exactly equal to the $\pi\pi$ phase shift, as one would think by a naive application of  Watson theorem. 
In the isobar  model, the complex constants multiplying the Breit-Wigner resonances describe the interaction of a resonance ($\sigma$) with  the third hadron ($\omega$). The $s$-dependent correction  $\Phi(s)-\delta_0^0(s)$ calculated above is generated by the individual interactions with $\omega$ of each of the outgoing pions.
Actually, the rescattering effect discussed above can be visualized by a triangular diagram, given in Fig. 1 of Ref. \cite{Anis1}. As shown in  \cite{Anis1}, this diagram is responsible for the appearance of anomalous singularities. In the present work we emphasized the influence of the rescattering effects on the phase of the $\sigma\to\pi\pi$ amplitude defined within the isobar model.

 We notice that,  if the total amplitude $A(s,t'')$  appearing in the r.h.s. of the unitarity relation (\ref{unit1}) could be expanded in a series of Legendre polynomials $P_l(\cos\theta_s)$, the standard evaluation of the integral \cite{MartinSpearman} would lead to Watson theorem for each partial wave $a_l(s)$.  In the case of the elastic $\pi\pi$ scattering (or in decays like $K\to\pi\pi l\nu$) such an expansion is legitimate, since in the physical region of the $s$-channel the amplitude is a holomorphic function of $t$. On the other hand, for three-body decays like (\ref{J}) a similar expansion is not possible, since  $P_l(\cos\theta_s)$, which are polynomials of $t$, fail to reproduce the branch cut along $t>(m_\pi+m_\omega)^2$. The isobar model attempts to take into account the singularities in all channels, but, as we discussed above, it is too simplistic. As a consequence, the phase of the $\sigma\to\pi\pi$ amplitude $a_0(s)$ defined in this model is not exactly equal to the phase-shift $\delta_0^0(s)$.

From (\ref{deltaPhi}) it follows that the magnitude of the phase difference $\Phi(s)-\delta_0^0(s)$  depends on the values of the parameters of the isobar model (the ratio $C_{b_1}/C_\sigma$ and the difference $\Delta_{b_1}-\Delta_\sigma$).  Since an overall constant phase is irrelevant, what  really matters is the variation with $s$. The difference $\Phi(s)-\delta_0^0(s)$ might be smaller than the experimental errors\footnote{For the decay $D^+\to\pi^+\pi^-\pi^+$, where the statistics is rather low, fits of equal quality were obtained both with  a phase of $a_0(s)$  very different from the $\pi\pi$ phase-shift \cite{E791}, and  with a phase close to $\delta_0^0$ \cite{Oller}.}. However, it is important  to emphasize that even a small phase difference may have an important influence on the pole determination. Indeed, an immediate consequence of our result is  that the denominator of the function $a_0(s)$ in (\ref{a0b0}) should not be identical to the denominator $D(s)$ appearing in  the expression (\ref{NoverD}) of $t_0^0(s)$. In Refs. \cite{BES, Bugg} the $\sigma$ pole is extracted from a parametrization of the denominator of $a_0(s)$ in the physical region, supposed to be valid also in the complex plane.  As mentioned in the Introduction, such a method is affected by the uncertainties of analytic continuation, which are large for a distant pole.  If, in addition, the denominator of $a_0(s)$  and the denominator of the $\pi\pi$ amplitude $t_0^0(s)$ differ in principle (even slightly) along the physical region, the position of the pole determined  by the analytic continuation of the  production data may be even more distorted.  The effect discussed above might play a role in understanding the difference between the mass and width of the lowest scalar resonance $\sigma$ extracted from the BES data for  $J/\psi\to\pi^+\pi^-\omega$ decay \cite{BES, Bugg}, and the values derived from Roy equation for the $I=l=0$ elastic $\pi\pi$ amplitude \cite{CCL}.

%%%%%%%%%%%%%%%%%%%%%%%%%%%%%%%%%%%%%%%%%%%%%
\vskip0.4cm
{\bf Acknowledgments}
\vskip0.2cm
 I am grateful to Prof. D. Bugg for drawing my attention  to previous papers on three-body decays, and to Prof. H. Leutwyler  for numerous discussions, comments and suggestions concerning the issues discussed above.
  This work is supported  by the Romanian MEdC under Contract CEEX 05-D11-49.

%%%%%%%%%%%%%%%%%


\begin{thebibliography}{99}
\bibitem{PDG 2004}
 S.~Eidelman {\it et al.}  [Particle Data Group],  Phys.\ Lett.\ B {\bf 592} (2004) 1.

\bibitem{CCL} I. Caprini, G. Colangelo and H. Leutwyler, Phys. Rev. Lett. {\bf 96} (2006) 132001.

\bibitem{BES} BES Collaboration (M. Ablikim et al.), Phys. Lett. B {\bf 598} (2004) 149.
\bibitem{E791} E791 Collaboration (E. M. Aitala et al.), Phys. Rev. Lett. {\bf 86} (2001) 770.
\bibitem{Bugg} D. V. Bugg, Eur.Phys.J.C {\bf 37} (2004) 433.

\bibitem{Oller} J. Oller, Phys. Rev. D{\bf 71} (2005) 054030.

\bibitem{Watson} K. M. Watson, Phys. Rev. {\bf 95} (1958) 316.


\bibitem{MartinSpearman} A.D. Martin and  T.D. Spearman, Elementary particle theory, North-Holland, 1970.

\bibitem{Anis1} 
 A.V. Anisovich and V.V. Anisovich, Phys. Lett. B345 (1995) 321.
\bibitem{Anis2}  A.V. Anisovich, Phys. Atom. Nucl. {\bf 66} (2003) 172.

 

\bibitem{LSZ} H. Lehmann, K. Symanzik,  W. Zimmermann, Nuovo Cimento,
{\bf 1} (1956) 205; {\bf 2} (1957) 425.


\bibitem{Bart} G. Barton,  Introduction to Dispersion Techniques in Field 
Theory, Benjamin, New York, 1965.


\end{thebibliography}
\end{document}